\documentclass{aes2e}

\usepackage{caption}
\usepackage{subcaption}
\usepackage{graphicx}
\usepackage{color} 
\usepackage{tabularx}
\usepackage{multirow}
\usepackage{booktabs}
\usepackage{interval}
\usepackage{ulem}
\usepackage{floatrow}
\usepackage{float}
\floatstyle{plaintop}
\restylefloat{table}

\jyear{xxx}
\jmonth{xxx}
\jvol{x}
\jnum{x}

\RequirePackage[2020-02-02]{latexrelease}
\begin{document}
	\markboth{Ackermann et. al}{Directivity of Musical Instruments}
	
	\title{A Database with Directivities of Musical Instruments 
		\thanks{Correspondence should be addressed to David Ackermann. E-mail: david.ackermann@tu-berlin.de}}
	
	\authorgroup{
		David Ackermann \quad\quad
		Fabian Brinkmann\quad\quad
		AND  \quad Stefan Weinzierl
		\affil{ Audio Communication Group, Technische Universität Berlin, Germany}
	}
	\abstract{
		We present a database of recordings and radiation patterns of individual notes for 41 modern and historical musical instruments, measured with a 32-channel spherical microphone array in anechoic conditions. In addition, directivities averaged in one-third octave bands have been calculated for each instrument, which are suitable for use in acoustic simulation and auralisation. The data are provided in SOFA format. Spatial upsampling of the directivities was performed based on spherical spline interpolation and converted to OpenDAFF and GLL format for use in room acoustic and electro-acoustic simulation software. For this purpose, a method is presented how these directivities can be referenced to a specific microphone position in order to achieve a physically correct auralisation without colouration. The data is available under the CC BY-SA 4.0 licence.
	}
	
	\maketitle
	
	\section{INTRODUCTION}
	Studies of the sound radiation characteristics of the human voice date back to the late 1930s \cite{dunnExplorationPressureField1939}, and studies of the directivity of musical instruments began thirty years later (summarized in~\cite[Chapter 4]{meyerAcousticsPerformanceMusic2009}). While these early measurements were often made with a single microphone moved around the source, the radiation patterns of acoustic sound sources such as speakers, singers, or musical instruments are today usually measured with the source at the center of an enclosing microphone array in anechoic conditions.
	
	Such a nearly full spherical array was used to analyze the directivity of 40 different human speakers, with measurements taken sequentially at 253 positions \cite{chuDetailedDirectivitySound2002}. The radiation characteristics of eight opera singers \cite{cabreraVocalDirectivityEight2007}  and fifteen trained singers \cite{monsonHorizontalDirectivityLow2012} were determined in the horizontal plane, with measurements taken at nine and thirteen positions, respectively. 
	High spatial resolution was employed for the measurements of a professional male singer, using an adjustable semi-circular microphone array with 24 receivers \cite{katzbrianDirectivityMeasurementsSinging2007}. A recent review of research on the sound radiation of singing voices is given by Abe (2019)~\cite{orieabeSoundRadiationSinging}.
	
	For the directivity of musical instruments, eight orchestral instruments were measured using 64 microphones \cite{hohlefabianKugelmikrofonarrayZurAbstrahlungsvermessung2009}, while 22 instruments of a symphony orchestra were measured using 22 microphones \cite{patynenDirectivitiesSymphonyOrchestra2010a}. A recently generated database for 14 instruments and a speaker contains radiation patterns measured at 2522 positions on a sphere \cite{leishmanViolinDirectivity2020}, but these data have only limited frequency resolution in (third-)octave bands. The most comprehensive and publicly available database was compiled for 41 modern and historic instruments, measured with 32 microphones, and includes recordings of single notes within the playable range of each instrument and directivities calculated from the stationary parts of these notes \cite{shabtaiGenerationAnalysisAcoustic2017, weinzierlADatabaseofAnechoicMicrophone2017}.
	
	Based on these measurements, the aim of this work was to provide the directivities of musical instruments in an open and standardised format, including both the acoustic measurements, the results of the subsequent processing, and important metadata, such as the exact position of the microphone capsules, the tuning frequency of the instrument and the pitch for which the directivity is valid. This should facilitate the exchange and use of such data by the scientific and general acoustics community. To facilitate this, we recently standardized the Spatially Oriented Format for Acoustics (SOFA) convention \texttt{FreeFieldDirectivityTF}, as part of the AES69-2022 standard \cite{aes_standards_comittee_aes69-2022, majdak_spatially_2022}, but also provide the data in OpenDAFF \cite{wefers2010free} and Generic Loudspeaker Library (GLL) \cite{AFMG} formats for use in room acoustic simulation software. 
	
	For each instrument, the database contains the single-tone recordings (calibrated to an absolute sound pressure and equalized for the microphone array transfer function), the extracted single-tone directivities, and the one-third octave band-averaged directivities and corresponding finite impulse responses (FIRs)\footnote{The data will be published after the review of this paper with a DOI on the DepositOnce repository and can be accessed beforehand at {https://tubcloud.tu-berlin.de/s/8joeeK3fFingLgp}}. 
	
	\section{METHODS}
	The directivities of 41 modern and historical musical instruments were measured using a 32-channel full-spherical microphone array in an anechoic chamber in an earlier study~\cite{shabtaiGenerationAnalysisAcoustic2017}. The following sections detail the recording of the original database and highlightes the improved processing of the data suggested in this work.
	
	\subsection{Data representation and format} \label{sec:back}
	The directivity of electro-acoustic sound sources, such as loudspeakers and microphones, can be described relatively easily by using transfer functions in the frequency domain or by finite impulse response (FIR) filters in the time domain for each direction of radiation. In contrast, the directivity of natural sound sources, such as human speakers, singers and musical instruments, is more complex to describe, because the directivity of a musical instrument depends not only on the frequency, but also on the note being played, and sometimes on the fingering, so that the same note can have multiple radiation patterns. 
	
	To ensure maximum flexibility, the \texttt{Freefield DirectivityTF} convention represents directivity data as complex transfer functions (TFs) at arbitrary frequencies. This allows FIR filters of artificial sound sources, but also multi-channel acoustic recordings of a musical instrument (see section \ref{sec:rec}) to be stored as complex spectra with linear frequency resolution. In addition, the directivity patterns of natural sound sources can be stored for arbitrary, not necessarily equidistant frequencies such as the fundamental frequency and the corresponding overtones (see section \ref{sec:tensor}), or as one-third octave frequency-band averaged data (see section \ref{sec:3rd_oct}). This representation is mainly used in geometric acoustic simulation. 
	
	For a complete description of natural sound source directivities, information about the instrument/singer and the way of playing is needed in addition to the measurement setup. The SOFA standard allows these data to be stored as metadata and uniquely assigned for easy handling and data exchange. An overview of the available metadata can be found in section \ref{sec:SOFA}.
	
	\subsection{Measurement setup}
	The instruments were recorded with a fully spherical lightweight microphone array in the anechoic chamber of the Technische Universität Berlin with a room volume of approximately $1070$~m$^3$ and a lower cut-off frequency of $f_{\text{c}} = 63$~Hz. 32 Sennheiser KE4-211-2 electret capsules of the microphone array were located at the vertices of a pentakis dodecahedron with a diameter of 2.1~m.
	
	A height-adjustable chair was used to position the musicians with their instruments so that the estimated acoustic centre of each instrument was as close as possible to the centre of the microphone array. The musicians faced the positive x-axis. The exact capsule positions of the array are given in spherical coordinates, i.e., in azimuth ($\phi = 0 ^\circ$ pointing in positive x-direction, $\phi = 90^\circ$ pointing in positive y-direction), colatitude ($\theta = 0^\circ$ pointing in positive x-direction, $\theta = 90^\circ$ pointing in positive z-direction) and distance (in metres) are included in the metadata of the SOFA files (cf. section \ref{sec:SOFA}). The recordings were made with 24 bit resolution and a sampling frequency of $f_s = 44.1$~kHz. The recordings were calibrated -- i.e., a digital amplitude of 1 corresponds to a pressure of 1 Pascal resp. a sound pressure level of $L_p = 94$~dB -- and compensated for the frequency response of the microphone array and the capsules. The exact measurement setup and a detailed description of the calibration procedure can be found in \cite{shabtaiGenerationAnalysisAcoustic2017}.
	
	\subsection{Processing}
	The calibrated and equalised single-tone recordings in the dynamic range of pianissimo (\textit{pp}) and fortissimo (\textit{ff}) provided as 32-channel WAV files, as published in~ \cite{weinzierlADatabaseofAnechoicMicrophone2017}, form the basis for the processing described below. 
	
	\subsubsection{Recordings} \label{sec:rec}
	The real-valued single-note recordings $x_q[n]$ of even length $N$ are available at discrete times $n\in \{0,1,...,N-1\}$ and for $Q=32$ channels of the spherical microphone array with $q\in \{1,2,...,Q\}$. Because the \texttt{FreeFieldDirectivityTF} convention requires frequency data, the recordings were Fourier transformed
	\begin{equation}
		X_q(k) = \sum_{n=0}^{N-1} x_q[n] \text{e}^{-\text{i} 2\pi \frac{k}{N}n},
	\end{equation}
	with $\text{i}^2=-1$ being the imaginary unit, and saved in the SOFA files as complex-valued single-sided spectra $X_{\text{S},q}(k)$ of length $N/2 + 1$
	\begin{equation}
		X_{\text{S},q}(k) = \left\{ 
		\begin{array}{l}
			X_q(k),  \quad \quad \ \text{if} \: k = 0 \\ 
			2 \cdot X_q(k), \quad \text{if } \: 0 < k < \frac{N}{2} \\
			X_q(k),  \quad \quad  \ \text{if} \: k = \frac{N}{2} .\\
		\end{array} \right.
	\end{equation}
	
	
	Note that reconstructing the time domain recordings from the published data thus requires the reconstruction of the both-sided spectrum of length $N$

	\begin{equation} \label{eq:sing2both}
		X_q(k) = \left\{ \begin{array}{l}
			X_{\text{S},q}(k),  \quad \quad \quad \quad \ \text{if} \: k = 0 \\ 
			\frac{1}{2} \cdot X_{\text{S},q}(k),\quad \quad \ \ \ \:  \text{if} \: 0 < k < \frac{N}{2} \\
			X_{\text{S},q}(k),  \quad \quad \quad \quad \ \text{if} \: k = \frac{N}{2}\\
			\frac{1}{2} \cdot X_{\text{S},q}^\ast(N-k),  \quad \text{if} \: k > \frac{N}{2}.
		\end{array} \right.\end{equation}

	with $(\cdot)^\ast$ denoting the complex conjugate, before applying the inverse Fourier transform (see ~\cite{Ahrens2020b} for details).
	
	\subsubsection{Single tone directivity data} \label{sec:tensor}
	To determine the directivity of the musical instruments, we used the stationary part of the single note recordings. For all instruments producing stationary parts, this was manually windowed by visual inspection, resulting in durations between 200 and 2104~ms, with a median duration of 630~ms. For the acoustic guitar and the harp, a quasi-stationary part was defined as the part between the decay time and the release time as estimated with the Timbre Toolbox \cite{pulkkiVirtualSoundSource1997}. For the transient timpani signals, the entire recording was used, from the onset to the transition into the noise floor.
	
\begin{figure}[tb]
		\centering
		\includegraphics[width=.49\textwidth]{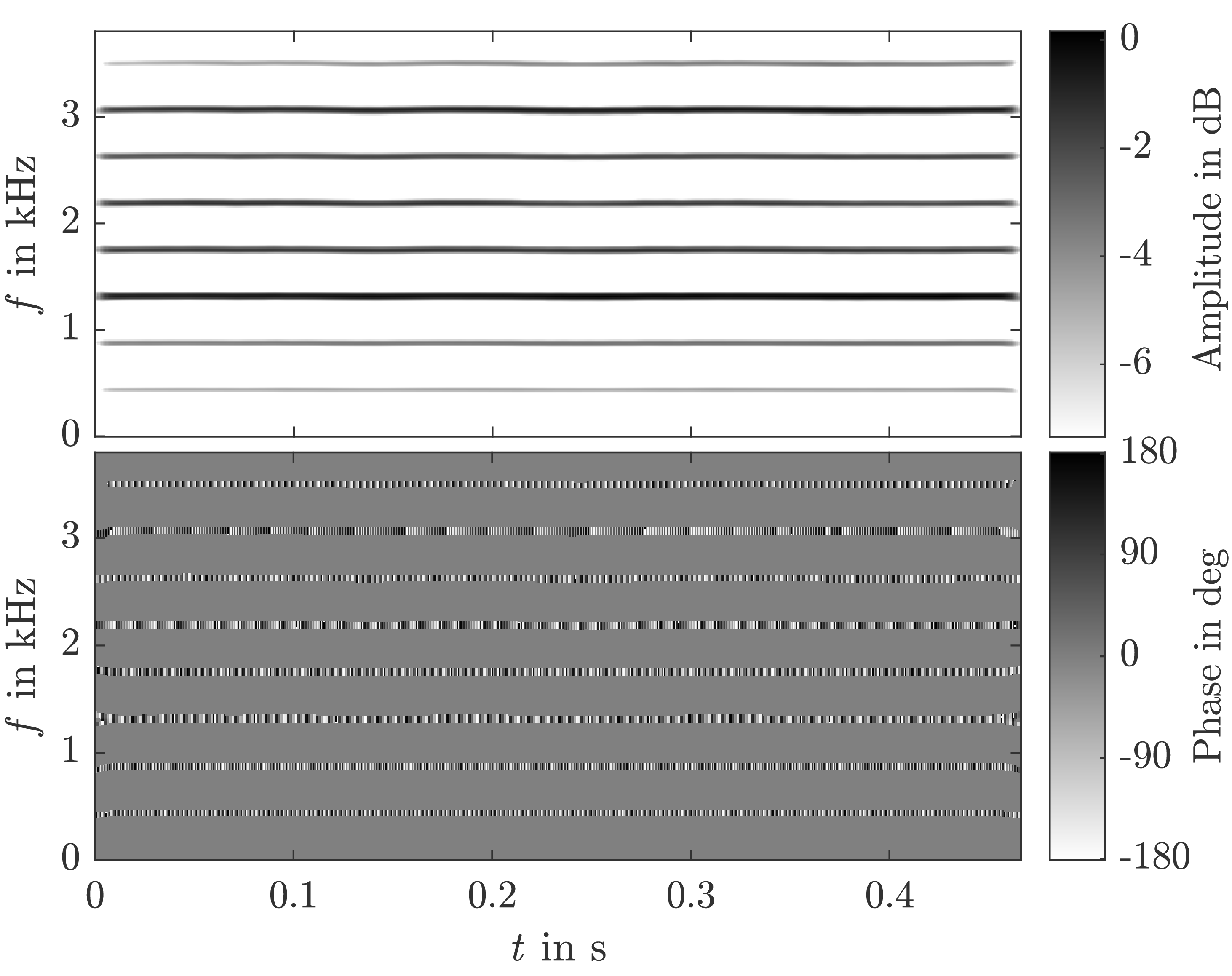}
		\caption{Extract from the spectrogram for the signal of a trumpet playing note A4 ($f_0 =440$~Hz) at fortissimo (microphone 4), with amplitude in dB (top) and phase in degree (bottom).}
		\label{phase}
	\end{figure}

	
	The directivities were estimated in two steps. First, the fundamental frequency $f_0$ and the frequency of the overtones $f_i$ in Hz were identified with $i\in \{1,2,...,I\}$ and $I$ being the highest identifiable overtone. Secondly, the energy at these frequencies was estimated. Both steps were based on the magnitude response $\left| X_q(k) \right|$ and ignored the phase information because natural sound sources, unlike artificial sound sources, do not have a stationary phase response, both for a given frequency and direction of radiation \cite{ahrensComputationSphericalHarmonic2021}. Figure \ref{phase} shows the magnitude and phase response for 0.5 seconds excerpt of a trumpet recording. Although the amplitudes of the fundamental and harmonics remain constant throughout the time window, the phase varies considerably, making it impossible to determine it unambiguously.
	
	This factor also complicates the use of natural sound sources for room acoustic simulation without further modification, since interpolation of the phase spectrum is highly susceptible to noise and can lead to errors, especially at high frequencies \cite{ahrensComputationSphericalHarmonics2020}. We have therefore proposed the use of absolute-valued directivities for interpolation~\cite{ackermann_comparative_2021}. This is in contrast to the previous processing of the data, which used the complex valued spectrum~\cite{shabtaiGenerationAnalysisAcoustic2017}.
	
	An estimate of the fundamental frequency $f_0$ was made by identifying the frequency with the highest amplitude within a window of $\pm 100$ cent bandwidth around the frequency corresponding to the tuning pitch indicated by the musicians. This was usually 442 or 443 Hz for modern instruments, 430 Hz for instruments of the Classical period, and 415 Hz for instruments of the Baroque period. This frequency was obtained for all 32 microphone recordings, and the most frequently occurring frequency over all 32 extracted values was chosen as $f_0$.
	
	In the next step, the frequencies of the partials were estimated by placing a search window of $\pm 10$ cents around each harmonic frequency corresponding to  $f_0$ and identifying the most frequently occurring, highest amplitude within this window, again considering all 32 microphone recordings. The search window takes into account the fact that, for physical reasons, the partials do not always lie exactly at the harmonic multiples of the fundamental frequency, and that the fundamental was not always held exactly constant over the duration of the stationary part. If the identified frequency deviated from the harmonic frequencies by more then five cents, the procedure was stopped, and the signal energy was considered to be below the noise floor. A visual inspection of the detected signal components confirmed that all relevant and clearly identifiable partial tones had been found by this procedure.
	
	At the frequencies $f_i$ that could be estimated, the directivity was calculated using the power spectral density (PSD)
	\begin{equation}
		S_{xx,q}(k) = \frac{1}{f_s N}|X_q(k)|^2\,
	\end{equation}
	which is a measure of the power of each frequency component in a signal with the implicit unit Pa\textsuperscript{2}/Hz.
	We have used the Welch method to improve the robustness of the PSD against noise by
	\begin{arabiclist}
		\item{}dividing the signal into eight segments of equal length with 50~\% overlap, 
		\item{}applying a Hanning window function to each segment to reduce the effect of spectral leakage caused by the finite length of the segments,
		\item{}computing the periodogram of each segment, which is an estimate of the PSD of that segment, and
		\item{}averaging the periodograms across the segments to obtain an estimate of the overall PSD of the signal.
	\end{arabiclist}
	To obtain an estimate of the power at each frequency, each estimate of the PSD was scaled by the equivalent noise bandwidth of the window (in Hz).
	
	\begin{figure*}[tp!]
		\floatbox[{\capbeside\thisfloatsetup{capbesideposition={right,top},capbesidewidth=4cm}}]{figure}[\FBwidth]
		{\caption{Normalized FFT (grey) and scaled PSD (black) of the signal of the modern oboe playing note C4 (262~Hz) at fortissimo, recorded with microphone 4. The asterisks indicate the detected power values and frequencies $f_0$ and $f_i$ that do not necessarily coincide with the discrete bins of the PSD.} \label{fig:directivity_computation}}
		{\includegraphics[width=.7\textwidth]{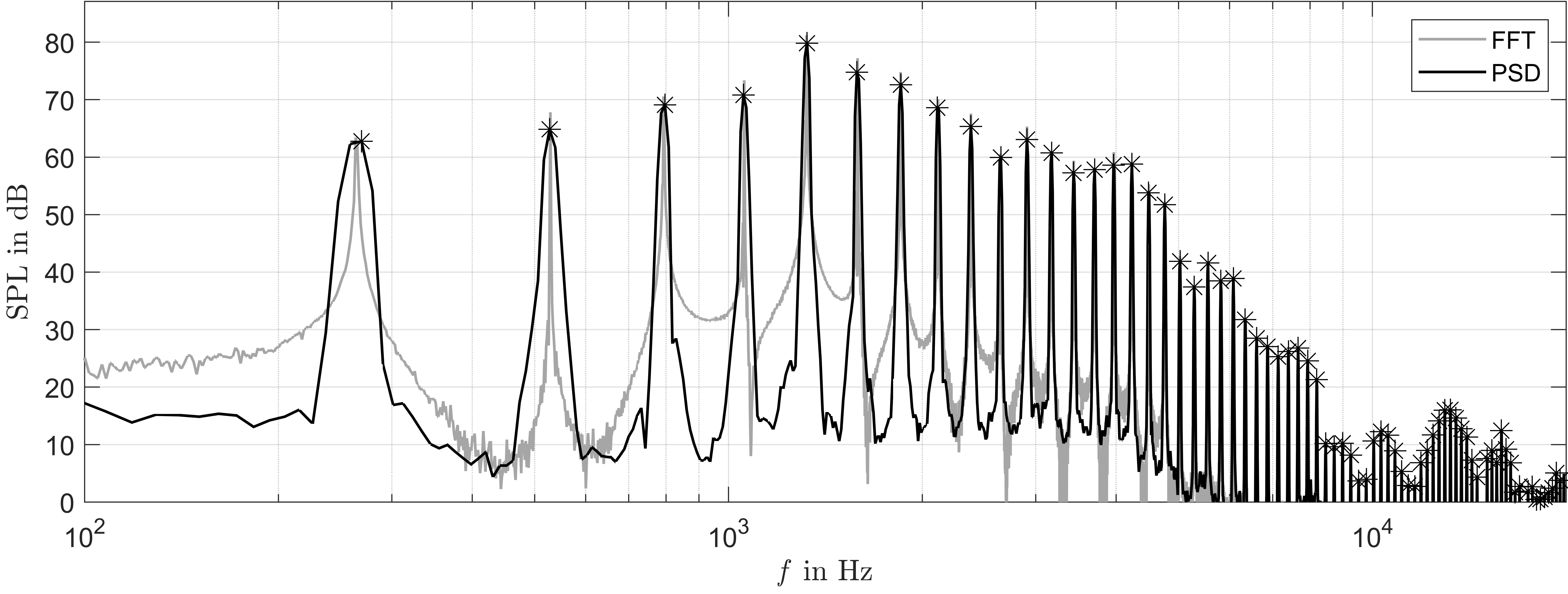}}
	\end{figure*}
	
	Finally, the power was determined by simple peak picking in the scaled PSD around the frequencies $f_i$ of the tone's harmonics and converted to the sound pressure value $p_{q}(f_i)$ by taking the square root. Figure \ref{fig:directivity_computation} illustrates the process for one note played by the modern oboe.
	
	\subsubsection{One-third octave band averaging} \label{sec:3rd_oct}
	Geometric acoustics applications typically require a single directivity pattern for a sound source, which is usually provided in one-third octave band resolution. 
	
\begin{figure}[h!]
		\centering
		\includegraphics[width=.47\textwidth]{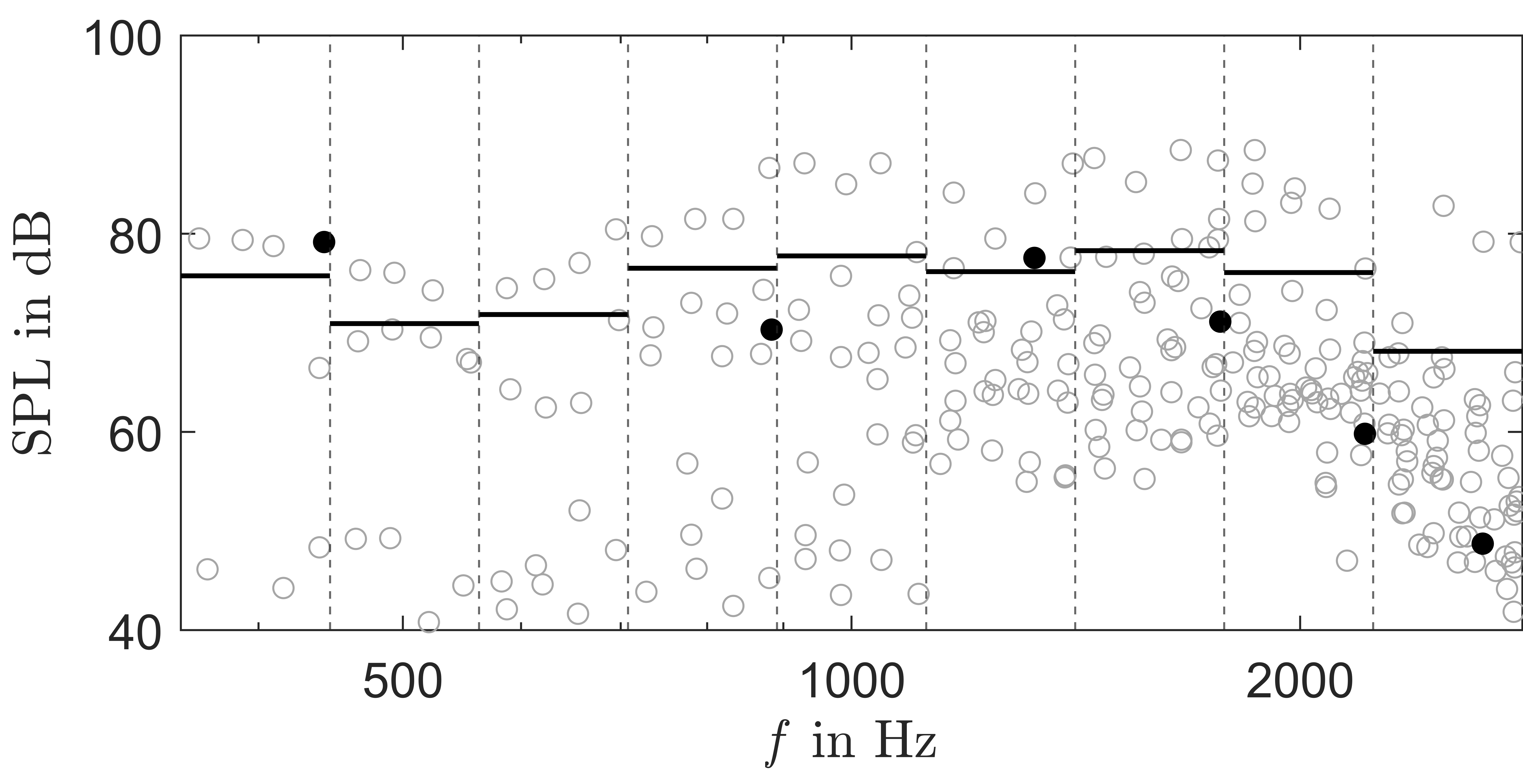}
		\caption{Estimated sound pressure levels of all partials of the modern oboe plotted against frequency (grey) in the one-third octave bands from 400 Hz to 2500 Hz, recorded with microphone no. 13. As an example, the fundamental and the first five overtones for the note A4 ($f_0=442$~Hz) are highlighted in black. The dashed vertical lines mark the boundaries of the third-octave bands. The horizontal solid black lines indicate the energetically averaged sound pressures $\bar p_{q,m}$ according to equation \ref{eq:3rd}, shown in dB.}
		\label{fig:3rd}
	\end{figure}

	This was calculated by energetic averaging of the partials of all individual notes of an instrument falling into the one-third octave frequency bands according to IEC 61260-1 \cite{IEC_61260}, for the $M=30$ centre frequencies from 25~Hz to 20~kHz. All partials of the $J$ individual notes $p_{q}(f_{i,j})$ from section \ref{sec:tensor} were used for the one-third octave band representation. The data were averaged for each of the $Q=32$ receivers of the spherical microphone array individually, by calculating the averaged amplitude as
	\begin{equation}\label{eq:3rd}
		\bar p_{q,m} = \sqrt{\frac{1}{L} \sum_{i=0}^{L-1} p_{q,i}^2},
	\end{equation}
	where $L$ indicates the number of partials identified in one one-third octave band.
	Figure \ref{fig:3rd} illustrates the averaging procedure over all partials for the modern oboe in one-third octave bands from 400~Hz to 2500~Hz.
	
	At this point, the data still contains the direction-dependent frequency response of each instrument. If the directivities are used for auralizations in which a simulated (binaural) room impulse response is convolved with an anechoic recording of an instrument, it should be noted that the anechoic recording also contains the frequency response of the instrument in the direction of the recording microphone. To avoid an unnatural coloration during auralizations, the frequency response has to be removed from the third-octave averaged directivities. In theory, this normalization should also consider the position of the microphone from which the anechoic recording of the instrument was made, i.e., this direction should be normalized to 0 dB. If this position, however, is unknown or unstable due to movements of the instrument relative to the microphone, it may be most robust to equalize the directivity so that equal energy is radiated across all directions within each band, i.e., to
	\begin{equation} \label{eq:p}
		\bar p_{\text{diff},q,m} = \frac{\bar p_{q,m}}{ \sqrt{ \sum_{q=1}^{Q} \bar p_{q,m}^2 \cdot w'_q}},
	\end{equation}
	where $w'_q$ are the normalized area weights of the measurement grid with $\sum w'_q = 1$.
	This representation of the radiation patterns is called \textit{diffuse} equalization in the following and will be discussed in more detail in section \ref{sec:disc}.
	
	The final step was to calibrate the directivity to the sound power of the real instruments, as the previous diffuse equalisation (cf. equation \ref{eq:p}) had lost the absolute sound power reference. This was done by averaging the sound pressure level over the effective one-third octave bands for each microphone. Third-octave bands with no sound radiation were set to zero. 
	
	The average sound pressure level of the diffuse equalized directivity per microphone was then calculated as
	\begin{equation}
		L_{\text{P},\text{3rd},q} = 10 \lg \left( \frac{\frac{1}{M}  \sum\limits_{m \in M}  \bar p^2_{\text{diff},q,m}}{p^2_0}  \right), 
	\end{equation}
	where $M$ indicates the number of effective one-third octave bands  and $p_0 = 2 \times 10^{-5}$ Pa. Its average over the surface of the spherical envelope is given by
	\begin{equation}\label{eq:spl}
		\bar L_{\text{P},\text{3rd}} = 10 \lg \left(\frac{1}{Q} \sum\limits_{q =1}^{Q}  10^{0.1  L_{\text{P},\text{3rd},q}}\right). 
	\end{equation}
	The sound pressure of the reference, i.e. the corresponding instruments, was calculated from the calibrated recordings as follows
	\begin{equation}
		L_{\text{P},\text{ref},q} = 10 \lg \left( \frac{\frac{1}{NJ} \sum\limits_{n= 1}^{N}  \sum\limits_{j= 1}^{J}  x^2_{q,j}[n]}{p^2_0}  \right), 
	\end{equation}
	where $N$ is the number of samples of the stationary part, $J$ is the number of single notes and averaged over all microphones to
	\begin{equation}\label{eq:spl}
		\bar L_{\text{P},\text{ref}} = 10 \lg \left(\frac{1}{Q}\sum\limits_{q =1}^{Q}  10^{0.1  L_{\text{P},\text{ref},q}}\right). 
	\end{equation}
	
	
	Finally, the calibration of the diffuse equalized directivity from equation \ref{eq:p} is given by
	\begin{equation}\label{eq:p_fin}
		\bar p_{\text{cal},q,m} =  \bar p_{\text{diff},q,m} \cdot 10^{\frac{\bar L_{\text{P},\text{ref}} - \bar L_{\text{P},\text{3rd}}}{20}}.
	\end{equation}
	The one-third octave band averaged directivity of each instrument is calculated for the dynamic fortissimo (\textit{ff}) and provided in the \texttt{FreeFieldDirectivityTF} SOFA convention.
	
	\subsubsection{Interpolation} \label{sec:interpolation}
	Several applications that rely on the directivity of sound sources, such as room acoustic simulations, require the use of continuous or high resolution data.
	Consequently, the measurement data must be spatially resampled (interpolated) to match the required sampling grid. 
	
	By sampling the actual sound pressure function $f(\theta,\phi)$ with a $Q$ channel spherical microphone array, the samples $p_q=f(\theta_q,\phi_q)$ are given at the positions $(\theta_q,\phi_q)$ of the respective microphones for $q\in \{1,2,...,Q\}$. The general mathematical formula for interpolation can therefore be expressed as
	\begin{equation}
		\hat f(\theta_r,\phi_r) = \sum\limits_{q =1}^{Q} f(\theta_q,\phi_q) \cdot L_q(\theta_r,\phi_r),
	\end{equation}
	where $\hat f(\theta_r,\phi_r) = \hat p_{r}$ is the estimated sound pressure at the $R$ points $(\theta_r,\phi_r)$ of the interpolation grid for $r\in \{1,2,...,R\}$ and $L_q(\theta_r,\phi_r)$  being the interpolation function derived from the known sound pressure $p_{q}$ at the position $(\theta_q,\phi_q)$. The specific choice of the interpolation function depends on the interpolation method being used.
	
	There is a plethora of techniques for interpolating real-valued scattered data that make different assumptions about the distribution of the discrete set of known data points \cite{liReviewSpatialInterpolation2008}. For musical instruments, the thin-plate pseudo-spline method \cite{wahbaSplineInterpolationSmoothing1981,wahbaErratumSplineInterpolation1982} of order 1 has been found to be a good method, producing lower interpolation errors than spherical harmonics (SH) interpolation \cite{weinreichMethodMeasuringAcoustic1980, zotterAnalysisSynthesisSoundRadiation2009, pollowDirectivityPatternsRoom2015a} or three-dimensional Vector-Based Amplitude Panning (VBAP~\cite{pulkkiVirtualSoundSource1997}) when applied to sparsely sampled directivity measurements and evaluated against the directivity of different musical instruments measured at high resolution as a reference~\cite{ackermann_comparative_2021}.
	
	We chose an equiangular grid with an angular resolution of $5^\circ$ in azimuth and colatitude as the target for the interpolation, resulting in $R = 2522$ sound pressure values $\hat p_{r,m}$ for each of the $m\in M$ one-third octave bands. The closed form spherical spline interpolation for order 1 was realized with AKtools using the function \texttt{AKsphSplineInterp()}~\cite{brinkmann2017aktoolsan} based on the directivity patterns according to equation \ref{eq:p_fin}.
	
	\begin{figure}[h!]
		\centering
		\begin{subfigure}[b]{0.20\textwidth}
			\centering
			\includegraphics[trim={.8cm .8cm 1.3cm .6cm},clip, width=\textwidth]{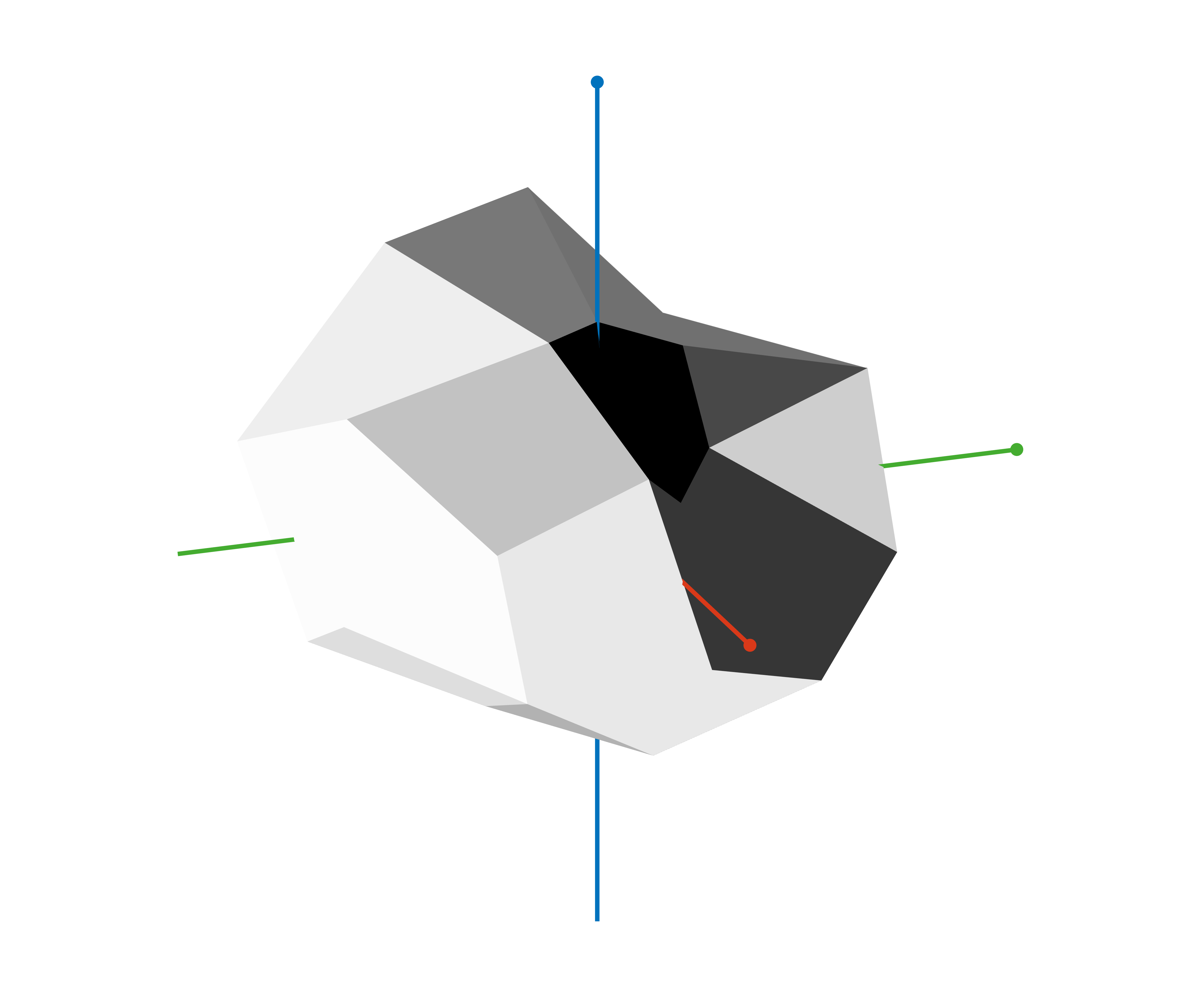}
			\caption{}
			\label{fig:y equals x}
		\end{subfigure}
		\hfill
		\begin{subfigure}[b]{0.20\textwidth}
			\centering
			\includegraphics[trim={.8cm .8cm 1.3cm .6cm},clip,width=\textwidth]{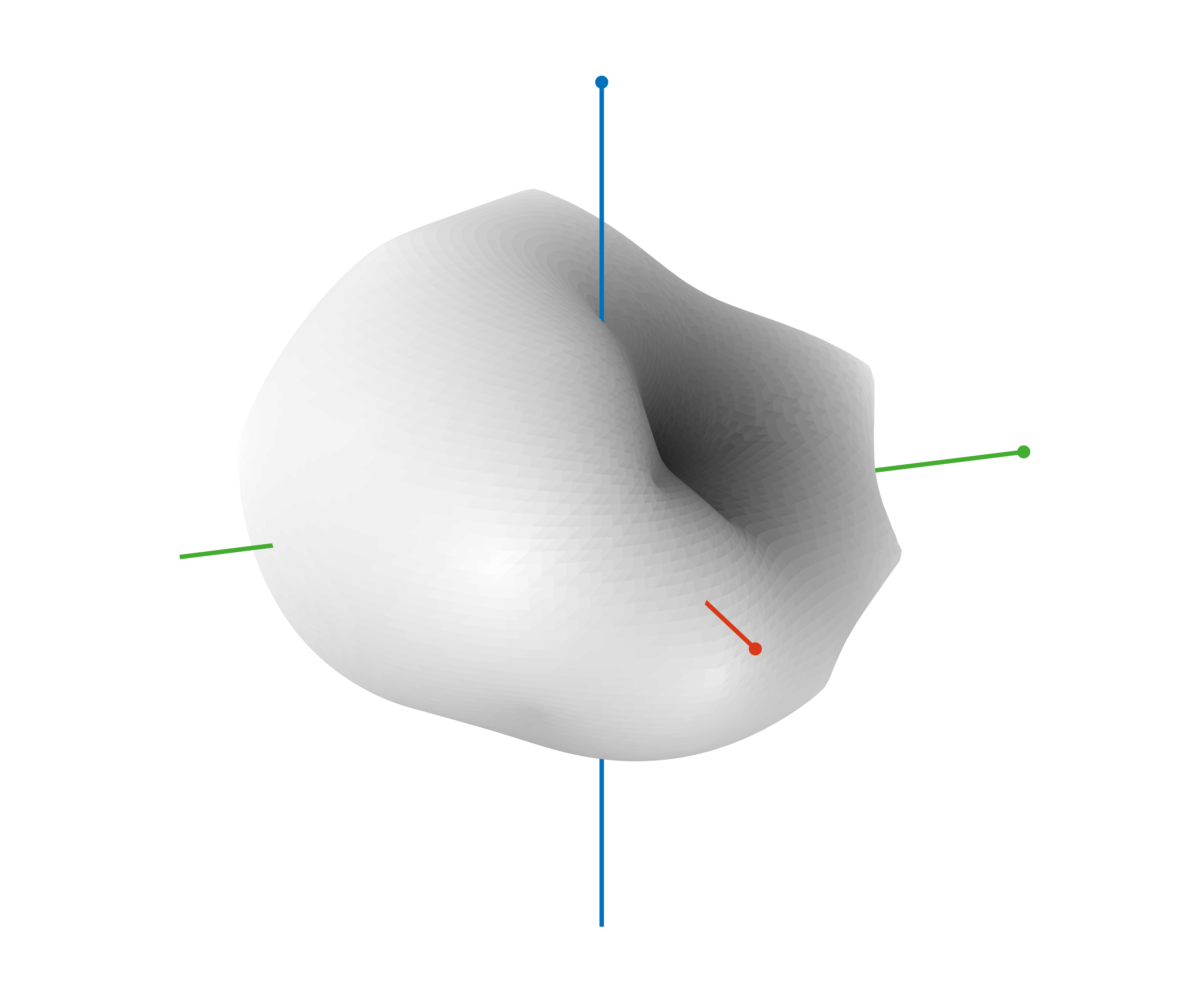}
			\caption{}
		\end{subfigure}
		\hfill
		\begin{subfigure}[b]{0.05\textwidth}
			\centering
			\includegraphics[scale=.35,trim={0cm 0cm 0cm 0cm},clip]{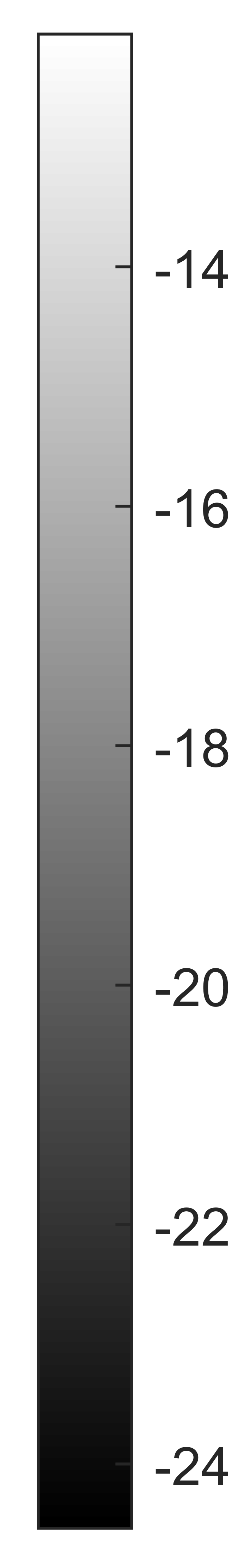}
			\caption*{}
		\end{subfigure}
		\caption{One-third octave band-averaged radiation pattern of the modern bassoon at 400~Hz, with original measurement resolution (32 points, a) and spatial interpolation to 2522 points, using spherical splines of 1st order (b). Balloon representation with radius corresponding to the sound pressure level in dB}.
		\label{fig:three graphs}
	\end{figure}
	
	\subsubsection{3rd octave smoothed FIRs} \label{sec:FIRs}
\begin{figure}[h!]
		\centering
		\begin{subfigure}[b]{0.47\textwidth}
			\centering
			\includegraphics[width=.98\textwidth]{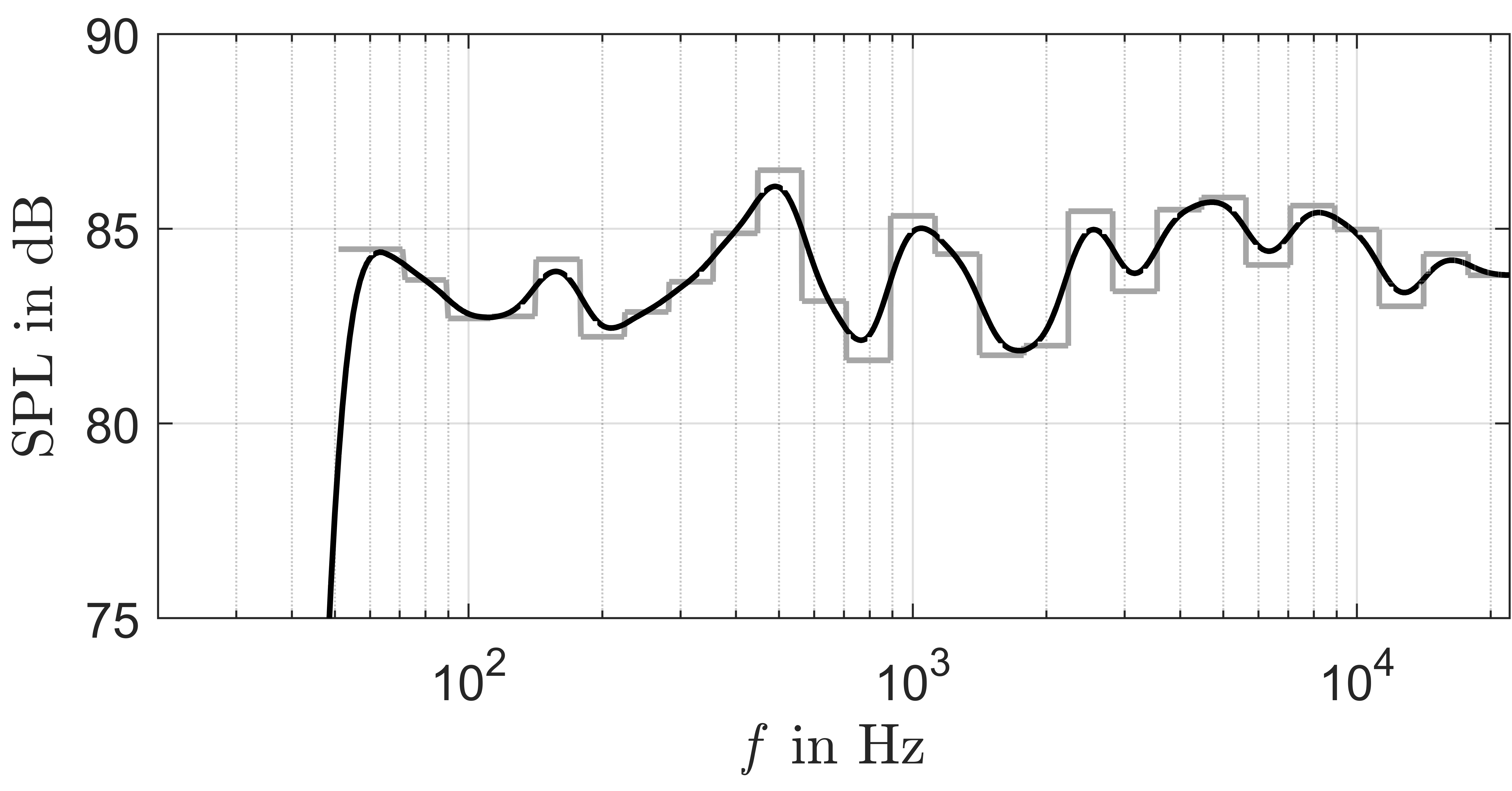}
			\caption{Magnitude spectrum}
			\label{fig:3rd_smooth}
		\end{subfigure}
		\hfill
		\begin{subfigure}[b]{0.47\textwidth}
			\centering
			\includegraphics[width=.98\textwidth]{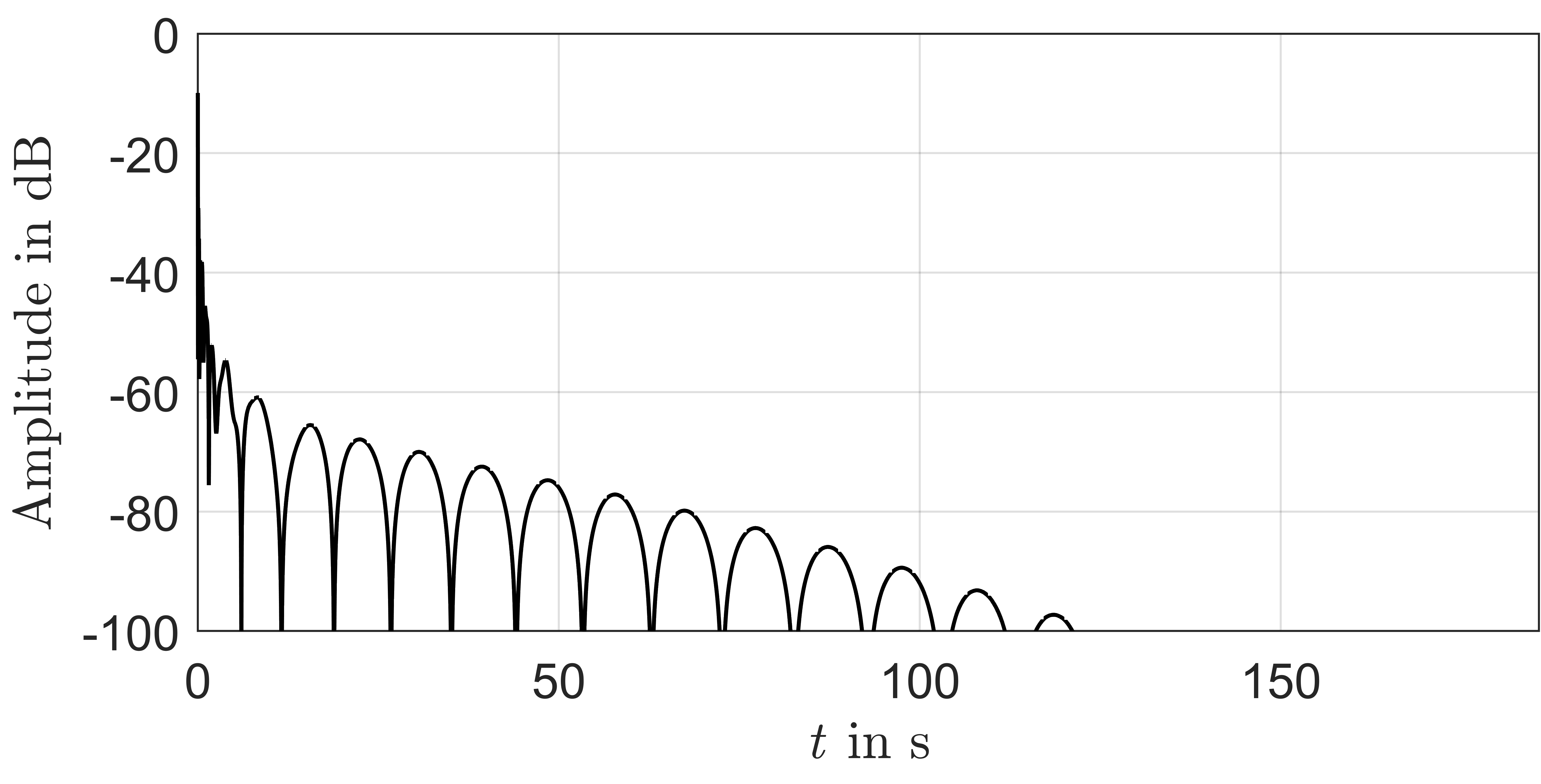}
			\caption{Energy Time Curve (ETC)}
			\label{fig:etc}
		\end{subfigure}
		\caption{Third-octave spectrum (gray) and smoothed third-octave spectrum (black) for the instrument bassoon modern (a) for the radiation direction $(\theta=0;\phi=-10)$ and the corresponding energy-time curve (b).}
		\label{fig:smooth}
	\end{figure}

	To enable musical instruments to be used as sound sources for simulating room acoustic and electroacoustic environments with software that uses FIR filters to represent directivity, $R = 2522$ FIR filters were calculated using the above mentioned grid. For this purpose, a one-third octave band spectrum according to IEC 61260-1 \cite{IEC_61260} with a frequency resolution of 1 Hz was generated from the estimated sound pressure values $\hat {p}_{r,m}$ (gray line in figure \ref{fig:3rd_smooth}). The one-sided spectrum with odd N was then smoothed with a one-third octave filter (black line in figure \ref{fig:3rd_smooth}). After transformation to a two-sided spectrum according to (equation \ref{eq:sing2both}) the spectrum was transformed into the time domain using IFFT and the phase was made minimum phase using the AKTools function \texttt{AKphaseManipulation()}. Finally, the FIR filter was reduced to 8192 samples according to AES56-2008 \cite{AES56-2008} as shown in figure \ref{fig:etc}.
	
	For use in the software EASE \footnote{www.afmg.eu/en (accessed May 2, 2023)}, this data was stored in the proprietary GLL format. 
	
	\section{DATABASE} \label{sec:SOFA}
	The database contains the calibrated single-note recordings, the single-note directivities, and the frequency-averaged directivity patterns in SOFA format \texttt{FreeFieldDirectivityTF} convention under a Creative Commons share alike licence (CC-BY-SA 4.0). In addition, high spatial resolution interpolated radiation patterns averaged over one-third octave bands, are provided in openDaff format and as FIR filters in GLL format. All data are freely accessible~\footnote{The data will be published after peer review with a DOI on the TU Berlin data repository and can be accessed beforehand at {https://tubcloud.tu-berlin.de/s/8joeeK3fFingLgp}}. A list of the available musical instruments can be found at table \ref{tab:instruments}.
	
	The SOFA files can be read using a variety of APIs\footnote{cf.  www.sofaconventions.org/mediawiki/index.php/\\Software\_and\_APIs (accessed May 2, 2023)}. They contain the recorded signals and extracted directivities as complex transfer functions (TFs) together with metadata describing the data in detail. This includes the name of the instrument in the entry \texttt{GLOBAL\_SourceName}, the name of the musician in \texttt{GLOBAL\_Musician}, the manufacturer of the instrument in \texttt{GLOBAL\_SourceManufacturer} and a verbal description of the position of the instrument during the measurement in \texttt{SourceView\_Reference}. The arrangement of the capsules of the microphone array is described in \texttt{ReceiverPosition}.
	
	In the SOFA data of the recordings, the respective note and the dynamic level are indicated in \texttt{GLOBAL\_Description}, e.g. \texttt{note = A4; dynamic = \textit{ff}}, the MIDI number in \texttt{MidiNote}, e.g. 69 for A4, and the frequency for A4 corresponds to the tuning frequency in the entry for \texttt{SourceTuningFrequency}. In the SOFA data of the original recordings, \texttt{SteadyPart} indicates the range of the manually determined stationary portion in samples. A detailed description of the structure of the database can be found in the accompanying documentation.
	
	\subsection{SOFA recordings}  \label{sec:SOFA_rec}
	The single note recordings are available as a one-sided complex TF for each instrument and note. The data can be converted into a two-sided spectrum according to equation \ref{eq:sing2both} and converted to the time domain by means of an IFFT. The naming of the data follows the scheme \texttt{SourceName\_dynamic\_midi Note\_recordings.sofa}, and the recordings are stored in the \texttt{Data.Real} and \texttt{Data.Imag} fields. The fields have the dimension \texttt{MxRxN}, where M (measurement) is always 1, R (receiver) is the number of capsules of the microphone array, with R = 32, and N indicates the length of the TF. The field \texttt{N} contains the frequencies of the bins of the TF in Hz.
	Note that the calibrated 32-channel WAV recordings on which this data set is based are still freely accessible~\cite{weinzierlADatabaseofAnechoicMicrophone2017}.
	
	\subsection{SOFA single note} \label{sec:SOFA_single}
	There is also a separate SOFA file for each instrument and note for the single-note directivity data; the naming of the file corresponds to the scheme \texttt{SourceName\_midiNote\_singleTones.sofa}. The purely real sound pressure levels are stored as complex transfer functions in the field \texttt{Data.Real} with the dimension \texttt{MxRxN}, where M is always 1, R = 32, and N refers to the number of the $I$ extracted partials. The \texttt{Data.Imag} field with the dimension \texttt{MxRxN} is included in the dataset for consistency reasons, but contains only zeros.The field \texttt{N} indicates the frequencies of the $I$ partials in Hz.
	
	\subsection{SOFA One-third octave band} 
	For the one-third octave band-averaged directivities, there is one SOFA file for each instrument with the naming scheme \texttt{SourceName\_3rdOctave.sofa}. The averaged and calibrated sound pressures from equation \ref{eq:p_fin} are stored in \texttt{Data.Real} with the dimension \texttt{MxRxN}, where M is always 1, R = 32, and N=30 refers to the nominal centre frequencies from 25~Hz to 20~kHz according to IEC 61260-1:2014 \cite{IEC_61260}. The data field \texttt{Data.Imag} has been filled with zeros. The field \texttt{N} indicates the centre frequencies of the one-third-octave bands in Hz. In this case, the entries \texttt{SourceTuningFrequency},  \texttt{SteadyPart} and \texttt{MidiNote} are not included in this data representation.
	
\begin{table}[h!]
	\label{tab:instruments} 
	\caption{List of modern (m) and historical (h) instruments whose individual directivities are available in the database in the indicated pitch range for the playing dynamics pianissimo (\textit{pp}) and fortissimo (\textit{ff}) and  additionally averaged in one-third octave bands.}
		\begin{tabular}{lccccc}
			Instrument 			& Era 	& pp 		& ff \\ \hline \hline
			Violin              & m     &  G3--G6 &  G3--G6  \\
			Violin              & h     &  G3--C7 &  G3--C7  \\
			Viola               & m     &  C3--F7 &  C3--F7  \\
			Viola               & h     &  C3--C6 &  C3--C6  \\[-.5ex]
			Cello               & m     &  C2--G6  & C2--G6  \\[-.5ex]
			Cello               & h     &  C2--D5 &  C2--D5  \\[-.5ex]
			Double bass         & m     &  E1--E5 &  E1--D\#5 \\[-.5ex]
			Double bass         & h     &  E1--E3 &  E1--E3  \\[-.5ex]
			Acoustic guitar     & m     &  C2--B5 &  E2--B5  \\[-.5ex]
			Double action harp  & m     &  C1--F4 &  C1--F4  \\[-.5ex]
			Oboe                & m     &  A\#3--A6 &  A\#3--F\#6  \\[-.5ex]
			Classic oboe        & h     &  C4--D\#6 &  C4--D\#6  \\[-.5ex]
			Romantic oboe       & h     &  B3--C6 &  D4--G5  \\[-.5ex]
			English horn        & m     &  E3--G\#5 &  E3--G\#5 \\[-.5ex]
			Clarinet            & m     &  D3--A\#6 &  D3--A\#6  \\[-.5ex]
			Clarinet            & h     &  D3--F6 &  D3--F6  \\[-.5ex]
			Bass clarinet       & m     &  A\#1--D\#5 &  A\#1--D\#5  \\[-.5ex]
			Bassoon             & m     &  A\#1--E5 &  A\#1--E5  \\[-.5ex]
			Classic bassoon     & h     &  A\#1--C5 &  A\#1--C5  \\[-.5ex]
			Baroque bassoon     & h     &  A1--F\#4 &  A1--F\#4  \\[-.5ex]
			Contrabassoon       & m     &  A\#0--D\#3 &  A\#0--F3  \\[-.5ex]
			Dulcian             & h     &  C2--G4 &  C2--G4  \\[-.5ex]
			Alto saxophone      & m     &  C\#2--C\#5 &  C\#2--C\#5  \\[-.5ex]
			Tenor saxophone     & m     &  B1--A4 &  B1--A4  \\[-.5ex]
			Flute   			& m     &  B3--A\#6 &  B3--D7  \\[-.5ex]
			Transverse flute 	& h     &  D4--F\#6 &  D4--A\#6  \\[-.5ex]
			Keyed flute         & h     &  C\#4--A6 &  C\#4--A6 \\[-.5ex]
			Trumpet             & m     &  F\#3--F6 &  F\#3--F6  \\[-.5ex]
			Natural trumpet     & h     &  D3--D6 &  D3--D6  \\[-.5ex]
			French horn         & m     &  D2--F\#5 &  D2--G5 \\[-.5ex]
			Natural horn        & h     &  A1--B4 &  A1--B4  \\[-.5ex]
			Basset horn         & h     &  F2--A\#5 &  F2--A\#5  \\[-.5ex]
			Tenor trombone      & m     &  G1--F5 &  G1--F5  \\[-.5ex]
			Alto trombone       & h     &  D2--D\#5 &  C\#2--D\#5 \\[-.5ex]
			Bass trombone       & m     &  B1--B4 &  C1--B4  \\[-.5ex]
			Tenor trombone      & h     &  E1--D5 &  E1--D5  \\[-.5ex]
			Tuba                & m     &  G\#0--C5 &  F0--E5 \\[-.5ex]
			Timpani             & m     &  D2  &  D2  \\[-.5ex]
			Pedal timpani       & m     &  D2 &  D2  \\[-.5ex]
			Soprano 			& --    &  G3--G\#5 &  G3--G\#5  \\[-.5ex]
	\end{tabular}
\end{table}
	
	\subsection{OpenDAFF One-third octave band} 
	The open source format openDAFF can be read with several APIs\footnote{cf. www.github.com/svn2github/opendaff (accessed May 2, 2023)}. The naming of the data follows the scheme \texttt{SourceName.daff}. The directional patterns are stored with a spatial resolution of $5^\circ$ (azimuth and colatitude), i.e. each file contains one-third octave magnitude spectra at 2522 points. These data can be used directly in the acoustic simulation environment RAVEN \cite{Vorlander2011a}. For evaluating the directivity, both of the individual notes and of the frequency-averaged directivity patterns, in arbitrary spatial resolution, we provide a Matlab script as part of the database (cf. section~\ref{sec:tools}).
	
	\subsection{GLL One-third octave band FIRs}
	The proprietary GLL format allows the integration of complex sound sources into the acoustic simulation environment EASE. The directional patterns averaged over a one-third-octave for all 21 modern musical instruments and for a soprano singer were stored as FIR filters with 8192 taps, and with a spatial resolution of $5^\circ$. For the exact naming scheme, please refer to the documentation of the dataset\footnote{The data will be published after the review of this paper with a DOI on the DepositOnce repository and can be accessed beforehand at https://tubcloud.tu-berlin.de/s/8joeeK3fFingLgp}. 
	
	The data can be converted to UNF (used by Ulysses\footnote{www.ifbsoft.de (accessed May 2, 2023)}) and the CLF/CIF format (used by ODEON \footnote{www.odeon.dk (accessed May 2, 2023)} and CATT-Acoustic\footnote{www.catt.se (accessed May 2, 2023)})  using the proprietary SpeakerLab\footnote{www.afmg.eu/en/ease-speakerlab (accessed May 2, 2023)} software from AFMG. 
	
	\subsection{Tools}\label{sec:tools}
	Part of the database is the \texttt{Directivity\_demo.m} Matlab script that makes it possible to read the recordings from the SOFA data, display their spectrum graphically and transform them into the time domain by IFFT. This data can then be saved as a WAV file and played back with common media players.
	
	The script also allows for the three-dimensional display of single-note and frequency-averaged directivity in the form of balloon plots based on the SOFA data provided. Finally, the data can be evaluated at any sampling quadrature using spherical spline interpolation. Prerequisite for all processing steps is the installation of AKTools\footnote{cf. www.tu.berlin/ak/forschung/publikationen/open-research-tools/aktools (accessed May 2, 2023)} and the SOFA API for Matlab (SOFAToolbox\footnote{cf. www.github.com/sofacoustics/SOFAtoolbox (accessed May 02, 2023)}) contained therein. 
	
	\section{DISCUSSION} \label{sec:disc}
	
	The present dataset contains recordings and radiation patterns of the individual notes of 41 modern and historical musical instruments, measured with a 32-channel microphone array in anechoic conditions. The recordings and directivities are provided in standardised SOFA format in convention \texttt{FreeFieldDirectivityTF}. From these data, averaged directivities have been calculated for each instrument, which are suitable for use in acoustical simulation and auralisation. In addition, spatially high-resolution directivities in OpenDAFF and GLL formats have been generated, allowing direct use in software such as RAVEN and EASE. 
	
	The absolute quality of the interpolation methods used for this spatial upsampling obviously depends on the characteristics of the sound source, such as its acoustically effective size, the modal patterns of its sound radiating parts, and the resulting complexity of the radiation pattern. For acoustically small sources, such as a trumpet and trombone bell or a violin, a fairly accurate interpolation can be expected even based on a measurement at 32 points. For extended sources with more complex radiation patterns, however, a sparse sampling grid may lead to increasingly poor estimates of the far field directivity \cite{ackermann_comparative_2021}. If other types of interpolation prove superior in the future, such as recently investigated, physically informed interpolation methods using the Euler equations as constraints~\cite{lemke2023}, the interpolations applied may be revised in the future. 
	
	For the physically correct auralisation of musical instruments in virtual acoustics, frequency-averaged directivities are a compromise that has to be made for technical reasons. Current simulation applications do not allow a straightforward exchange of single tone representations of such directivities within a simulation run. Due to the frequency averaging of the input data, a tonal colouration of the simulation result may occur \cite{corcuera2023perceptual}. The magnitude of this effect and its influence on instrumental and room acoustic perception will have to be determined in a subsequent study using the data presented.
	
	When an anechoic recording of an instrument and its directivity is used to auralize a virtual acoustic environment, it is essential to normalize the directivity of the source to a suitable reference. In theory, this would be the position of the microphone used to make the anechoic recording. Since the recorded signal already contains the directional timbre characteristic of the instrument in that direction, it should not be altered by the applied directivity, as this would result in unwanted colouration of the simulation.
	This means that the directivity should be normalized, so that only a frequency response \textit{relative} to this reference direction will be obtained. This will be achieved by calculating 
	\begin{equation}
		\hat p_{\text{pt},r,m} = \frac{\hat p_{r,m}}{ \hat p_{\text{mic},m}},
	\end{equation}
	where $\hat p_{\text{mic},m}$ is the interpolated and frequency-averaged sound pressure of the $M$ one-third octave band in the reference direction.
	
	In a real recording situation, however, the instrument being played by a musician is a moving sound source. This can result in an angular displacement of the instrument which can easily reach up to $47^\circ$ when played in a standing position and up to $36^\circ$ when played in a sitting position \cite{ackermannAcousticalEffectMusicians2019}. To compensate for the movement of the instrument when referencing, we consider it useful to equalise the directivity not to a point, but to the average of a larger spherical surface area $A$. The distribution of orientations over this surface can be taken into account by using a weighting function, i.e., by calculating 
	
	\begin{equation}
		\hat p_{\text{area},r,m} = \frac{\hat p_{r,m}}{ \sqrt{ \sum\limits_{r_A=1}^{R_A} \hat p_{r_A,m}^2 \cdot w'_{r_A} \cdot g'_{r_A}}},
	\end{equation}
	where $w'_{r_A}$ and $g'_{r_A}$ are the normalized area and the two-dimensional function weights, respectively, with
	\begin{equation}
		\sum\limits_{r_A=1}^{R_A} w'_{r_A} \cdot g'_{r_A} = 1,
	\end{equation}
	and  $r_A\in \{1,2,...,R_A\}$ are the $R_A$ grid points of the area $A$ over which the mean is to be calculated.
	
	If no information about the position of the microphone during the recording is documented, a diffuse equalised directivity (cf. equation \ref{eq:p}) can be used to minimise the sound colouration on average. For this reason, the directivity averaged over one-third octave bands has been provided in this representation for uniformed immediate use. 
	
	The extent to which the different referencing methods of the recording position affect the perceived sound event and the room acoustic parameters will be clarified in a later investigation on the basis of these data.
	
	\section{ACKNOWLEDGMENT}
	We would like to thank Stefan Feistel and Silke Bögelein (AFMG) for their support in producing the GLL datasets.
	\bibliography{bibliography.bib}
	\bibliographystyle{aes2e.bst}

\end{document}